\documentclass[lettersize,journal]{IEEEtran}
\usepackage{amsmath,amsfonts}
\usepackage{algorithmic}
\usepackage{array}
\usepackage[caption=false,font=normalsize,labelfont=sf,textfont=sf]{subfig}
\usepackage{textcomp}
\usepackage{stfloats}
\usepackage{url}
\usepackage{verbatim}
\usepackage{graphicx}
\usepackage{tabularx}
\usepackage{booktabs}
\usepackage[sorting=none]{biblatex}

\def\BibTeX{{\rm B\kern-.05em{\sc i\kern-.025em b}\kern-.08em
    T\kern-.1667em\lower.7ex\hbox{E}\kern-.125emX}}
\usepackage{balance}

\newcommand{\ie}{i.\,e.}
\newcommand{\eg}{e.\,g.}

\begin{document}
\title{Estimating ensemble likelihoods for the Sentinel-1 based Global Flood Monitoring product of the Copernicus Emergency Management Service}
\author{Christian Krullikowski, Candace Chow, Marc Wieland, Sandro Martinis, Bernhard Bauer-Marschallinger, Florian
Roth, Patrick Matgen, Marco Chini \IEEEmembership{Senior Member, IEEE}, Renaud Hostache, Yu Li, Peter Salamon

\thanks{This work was supported by the European Commission through the project "Provision of an Automated, Global, Satellite-based Flood Monitoring Product for the Copernicus Emergency Management Service" (GFM) under Grant JRC/IPR/2020/OP/0551. Corresponding author: Christian Krullikowski. The authors Christian Krullikowski, Marc Wieland and Sandro Martinis are and the author Candace Chow was with the German Remote Sensing Data Center (DFD), German Aerospace Center (DLR), D-82234 Oberpfaffenhofen, Germany (email: christian.krullikowski@dlr.de; marc.wieland@dlr.de; sandro.martinis@dlr.de; c.chow@alumni.utoronto.ca).

The authors Bernhard Bauer-Marschallinger and Florian Roth are with the Remote Sensing Research Group, Department of Geodesy and Geoinformation, TU Wien, 1040 Vienna, Austria (email: bbm@geo.tuwien.ac.at; florian.roth@geo.tuwien.ac.at). The authors Patrick Matgen, Marco Chini and Yu Li are and Renaud Hostache was with the Environmental Research and Innovation Department, Luxembourg Institute of Science and Technology, Esch-sur-Alzette, Luxembourg (email: patrick.matgen@list.lu; marco.chini@list.lu; yu.li@list.lu). The author Peter Salamon is with the Joint Research Centre (JRC) of the European Commission, Ispra, Italy (email: peter.salamon@ec.europa.eu).}}

\maketitle

\begin{abstract}
The Global Flood Monitoring (GFM) system of the Copernicus Emergency Management Service (CEMS) addresses the challenges and impacts that are caused by flooding. The GFM system provides global, near-real time flood extent masks for each newly acquired Sentinel-1 Interferometric Wide Swath Synthetic Aperture Radar (SAR) image, as well as flood information from the whole Sentinel-1 archive from 2015 on. The GFM flood extent is an ensemble product based on a combination of three independently developed flood mapping algorithms that individually derive the flood information from Sentinel-1 data. Each flood algorithm also provides classification uncertainty information that is aggregated into the GFM ensemble likelihood product as the mean of the individual classification likelihoods. As the flood detection algorithms derive uncertainty information with different methods, the value range of the three input likelihoods must be harmonized to a range from low [0] to high [100] flood likelihood. The ensemble likelihood is evaluated on two test sites in Myanmar and Somalia, showcasing the performance during an actual flood event and an area with challenging conditions for SAR-based flood detection. The Myanmar use case demonstrates the robustness if flood detections in the ensemble step disagree and how that information is communicated to the end-user. The Somalia use case demonstrates a setting where misclassifications are likely, how the ensemble process mitigates false detections and how the flood likelihoods can be interpreted to use such results with adequate caution.
\end{abstract}


\section{Introduction}
\IEEEPARstart{W}{ith} an amount of 44 \% \cite{DisarmamentAffairs2020} of all occurring disasters and produced economic losses of about 651 billion dollars, floods are among the most severe disasters worldwide. Although not the deadliest natural disaster, floods are affecting the largest number of people worldwide every year. With globally rising temperatures, Dottori et al. (2018) \cite{Dottori2018} predict an increase in human losses due to flooding by up to 70 to 83 \% and additional direct flood damages upwards of 160 to 240 \%. Botzen et al. (2019) \cite{Botzen2019} identify population and economic growth in disaster-prone regions as key causes leading to this increase. Apart from human losses, floods may cause damages to (critical) infrastructure \cite{Kadri2014} and may lead to further cascading effects such as the spread of infectious diseases \cite{Suk2019}. 

Mitigating these effects requires coordinated action on multiple levels, including but not limited to the implementation of accurate early warning systems, constant monitoring of disaster-prone regions and well implemented risk management procedures \cite{Sendai2015}. Arguably, the monitoring requirement, especially at large scale, is currently best fulfilled through the utilization of Earth Observation data. Grimaldi et al. (2016) \cite{Grimaldi2016} present a review of different flood data sources and compare optical with synthetic aperture radar (SAR) sensors. Optical imagery mainly relies on cloud-free and illuminated data, whereas radar remote sensing satellites can operate day and night due to their ability to emit cloud penetrating microwave. 

Past studies already highlighted the potential of a synergetic use of optical and SAR data in flood mapping \cite{Chaouch2012} and \cite{Spasova2019}. However, most studies focus on a single technology, most frequently microwave remote sensing \cite{Clement2018}, \cite{Jo2018}, \cite{Pulvirenti2011} and \cite{Huang2018}. A comprehensive overview of advantages and limitations of different methods is found in \cite{Schumann2015}. 

It can be noted that the aforementioned studies either focus on specific regions or were not implemented as operational services. Furthermore, a majority of the studies do not provide any information on the flood classification uncertainties. Clement et al. (2018) \cite{Clement2018} highlight several sources of uncertainty affecting SAR-based flood extent mapping, for example the ambiguities related to similar backscatter return over water look-alikes and dry soil, as well as areas with fuzzy backscatter response, \eg \space dense vegetation, and areas with higher backscatter return over urban areas. In general, cases where SAR-based flood mapping may be hampered and the detection may become less confident, the classification necessitates and benefits from the inclusion of a dedicated uncertainty analysis \cite{Hertel2023}. This complementary output also supports the interpretation and use of SAR-based flood map products, where end-users can be alerted to flood features associated with lower confidence, which should be treated with more caution with respect to risk assessment. 

The absence of a fully operational flood service that also returns confidence information culminated in the request of the Copernicus Emergency Management Service (CEMS) to integrate technically mature and scientifically validated flood detection algorithms into the Global Flood Awareness System (GloFAS) \footnote{https://www.globalfloods.eu/} and the European Flood Awareness System (EFAS) \footnote{https://www.efas.eu/}. Instead of utilizing an approach based on a single retrieval algorithm, the Joint Research Centre (JRC) as the contracting authority adopted an ensemble approach, which merges the results of three matured and independently developed flood algorithms. 

\subsection{Conceptual basis of GFM ensemble}
\noindent The Global Flood Monitoring (GFM) product of the CEMS continuously processes and analyzes all incoming Sentinel-1 Ground Range Detected (GRD) Interferometric Wide swath (IW) data, aiming to detect and monitor flood events in nearreal time at global scale. 

The GFM product builds on an ensemble approach that combines three mature and independently developed flood detection algorithms provided by the German Aerospace Center (DLR), Luxembourg Institute of Science and Technology (LIST) and Vienna University of Technology (TUW). The flood ensemble is computed pixel-wise and based on a majority voting system, where at least two algorithms must classify a pixel as flooded or non-flooded. Further insight into the flood ensemble algorithm is described by \cite{Chini2023}. 

Besides a pixel-based flood classification, each flood detection algorithm generates classification uncertainty information in the form of likelihoods. The GFM ensemble algorithm then combines the three individual layers of uncertainty information into a single layer termed ensemble likelihood. Although, an uncertainty analysis is performed, the term likelihood is used instead of uncertainty; most users have an a-priori understanding about likelihoods, whereas uncertainties describe a negation which may not be understood as intuitively. 

The GFM product is composed of different layers supporting the interpretation of flood situations using remote sensing data. Besides the actual flood extent layer, users can also download the likelihood data from GloFAS and EFAS. In addition, a downloadable exclusion layer informs about regions where no flood delineation was possible. These areas correspond to nodata values in both of the flood and likelihood products.

\subsection{Objectives of ensemble flood detection and interpretation of
ensemble likelihoods}
\noindent Two sets of objectives drive the ensemble-based flood detection and the manner in which ensemble likelihoods are intended to be interpreted and applied by two user communities: 1) integrating results into further processes or studies and 2) utilizing results for decision-making processes. 

The first objective addresses the first community, consisting of algorithm developers. Ensemble likelihoods may be used by them to identify subsets of pixels associated with low confidence values as a way to gain insights about opportunities for improving algorithms so that they return more accurate predictions. The individual and combined likelihoods may also serve as a basis for inter-comparing the results obtained by different approaches, thereby potentially improving our understanding of their strengths and weaknesses. 

The second community are data (end-)users. They may use ensemble likelihoods to minimize adverse consequences of making potentially costly decisions based on highly uncertain information. Results of this study may thus provide a basis on how the two aforementioned use cases can be used to support decision-making for flood and non-flood events. 

In particular, to evaluate the impact of differential decisions made based on flood classifications with and without consideration of ensemble likelihood values, we examine two land cover types (\ie \space agricultural lands, built environments) of particularly high economic importance and social consequences. In effect, mean likelihoods serve as a heuristic indication of overall confidence in the flood prediction, based on an average of available flood and respective uncertainty outputs from contributing individual flood algorithms. The values also function as an indicator of the current capacities/confidence of ensemble algorithms to detect water over certain types of land covers and uses. 

This study focuses on gaining insights on scenarios that result in regular and over-detections with respect to dominant land covers. Based on the results, benefits and limitations of the ensemble likelihood approach are highlighted and provide a starting point to guide further developments and applications in the two communities. 

The application of ensemble likelihoods is evaluated with two use cases exemplifying flood (Myanmar use case) and non-flood (Somalia use case) events, respectively. In particular, the objectives are to accurately delineate flood extent, while minimizing over- and under-detection. Extension of case-based assessments provide useful insights on the generalizability of the flood monitoring algorithm on a global scale with respect to a more comprehensive range of land covers/ uses. 

Subsequent sections provide detailed descriptions about how each individual flood algorithm generates uncertainty information (Section 2), the generation and evaluation of ensemble likelihoods (Section 3). Data used to conduct the study is described in Section 4, followed by results (Section 5), discussion (Section 6), conclusions and recommendations (Section 7).

\section{Generation of algorithm likelihoods}
\noindent The GFM flood ensemble likelihood product attributes to each valid pixel a likelihood of being flooded given its recorded Sentinel-1 backscatter value and ancillary data inputs. The term valid refers to pixels that are considered to be potentially flooded and included in the computation. Invalid pixels are excluded through an exclusion mask. This mask excluded areas blocked by radar shadow, regions of no Sentinel-1 SAR sensitivity towards flood dynamics, or areas that are considered non-floodable as they are located too far away from the next drainage \cite{Nobre2011}. 

\begin{figure}[!t]
	\centering
	\includegraphics{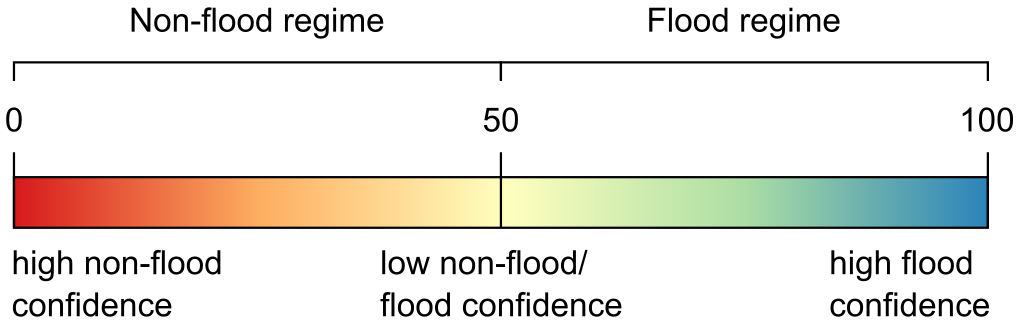}
	\caption{Confidence distribution of likelihood values. Likelihood values towards 0 correspond to higher confidence in non-flood classifications, whereas values towards 100 correspond to higher confidence in flood classifications. Low confidence in both classifications is indicated by likelihood values towards 50.}
	\label{fig:likelihoodrange}
\end{figure}

Ensemble likelihoods are defined in the interval [0, 100]. Likelihood values towards 0 represent lowest confidence in the ensemble flood classification, whereas values towards 100 represent highest confidence. Ensemble likelihoods are used to convert the set of ensemble classifications into a single binary flood classification, representing non-flood pixels as 0 and flood pixels as 1, respectively. In this binarization step, a likelihood value of 50 is defined as the threshold value that separates the two classes (\ie, non-flood pixels from the interval [0, 49] and flood pixels from the interval [50, 100]). Confidence about the detection of each respective class increases with likelihood values towards the lower or higher class boundaries (see Fig. \ref{fig:likelihoodrange}). 

The ensemble likelihood value is computed pixel-wise as the mean of the likelihood values attributed to each valid pixel by the three algorithms. The following sections describe the independent generation of each set of values. 

In general, all likelihoods are in reference to a flood classification. If the likelihood is low over a certain pixel or feature, the classification confidence that the pixel or feature is flooded is also low. The specific terms that are further evaluated in the next sections are defined as follows:

\begin{itemize}
	\item{individual likelihood values: refer to pixel-wise likelihood information generated by each of the three flood algorithms (\ie, DLR, LIST, TUW)}
	\item{initial mean likelihood values: are computed pixel-wise based on the average of all available individual likelihood information, ideally generated by all three flood algorithms, prior to the application of the ensemble algorithm}
	\item{ensemble likelihood values: are updated likelihood values based on the initial mean likelihood after the ensemble algorithm is applied}
\end{itemize}

The ensemble algorithm combines the flood detection and likelihood outputs of the individual flood algorithms. Although a majority voting system is implemented, split situations, \ie, cases of classification disagreement occur where a majority cannot be achieved, \eg \space when one out of three algorithms yields a nodata pixel. 

Post-processing steps involve the exclusion of sub-areas within a given Sentinel-1 scene that overlap with the reference water and exclusion masks. Clusters with a size less than a defined threshold of flood pixels are assumed to be unlikely flooded and re-labeled as non-flood pixels. This action is termed a blob removal step and eliminates small fragmented patches. Results following these steps are then referred to as ensemble classifications, see \cite{Chini2023}. Likelihood values corresponding to formerly flooded but excluded pixels are set to a likelihood value of 0. Likelihood values corresponding to formerly flooded but blob-removed pixels are set to a likelihood value of 49, \ie, expressing the lowest confidence in the non-flood regime. 

The ensemble algorithm can be applied based on two approaches: split and consensus. The split approach considers the likelihood values associated with the respective classification of each flood algorithm and favors the classification with the highest confidence. The consensus approach is based on majority voting, which sets all split situations to non-flooded classifications, since only 2/3 flood algorithms generate valid but conflicting pixel-wise classifications. In effect, a flood classification is only returned when there is an agreement. The following sub-sections describe the individual likelihood layers produced by the individual flood detection algorithms.

\subsection{Computation of DLR fuzzy values}
\noindent The flood detection algorithm by DLR is a single scene approach, \ie, the main data input for flood inundation is a single Sentinel-1 observation. The DLR algorithm applies fuzzy logic post-processing to measure and to reduce the uncertainty associated with the water classification, originally described by \cite{Martinis2015} and \cite{Twele2016}. Three cases influence classification uncertainty. In particular, the likelihood of a pixel being classified as water is low:

\begin{itemize}
	\item{if it's radar backscatter is close to the automatically derived threshold $\tau$, separating water and non-water;}
	\item{if the slope at that location is high, since steeper surfaces are unlikely to retain water; and}
	\item{if that pixel is connected to other neighboring water pixels and the resulting area is relatively small. On the contrary, the uncertainty is low if the pixel is connected to other neighboring water pixels and the resulting area is relatively large.}
\end{itemize}

The three cases or parameters, namely backscatter of the normalized radar cross section (NRCS), slope and minimum mapping unit are evaluated separately, resulting in the generation of three fuzzy layers. The concept of the fuzzy logic step is exemplified with the consideration of SAR backscatter values. 

\begin{figure}[!t]
	\centering
	\includegraphics{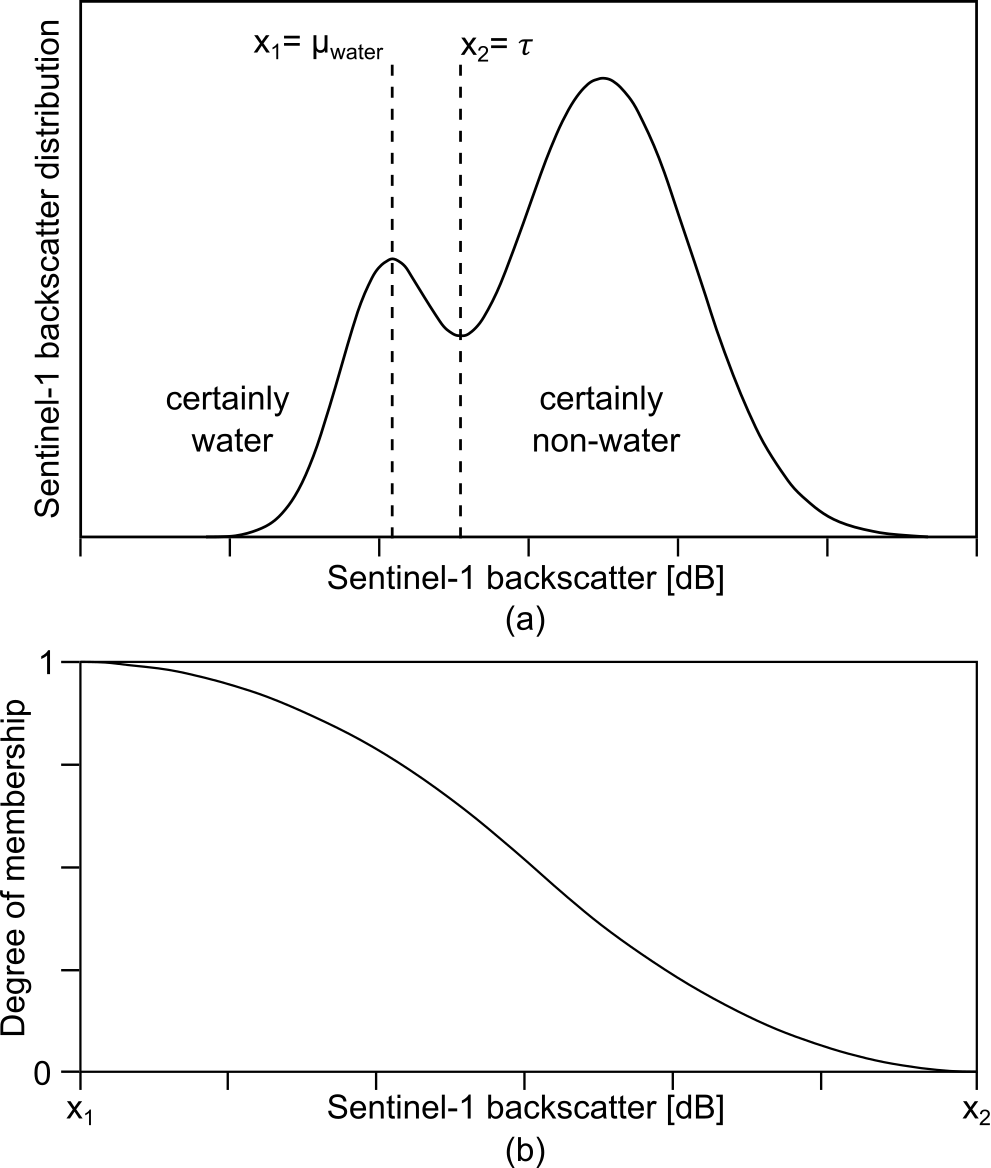}
	\caption{Fuzzy logic approach for discriminating between water
and non-water classes based on SAR backscatter values,
denoted as $\sigma^0$ Sentinel-1 backscatter [dB].}
	\label{fig:dlr-fuzzy}
\end{figure}

Fig. \ref{fig:dlr-fuzzy} illustrates the application of the fuzzy logic approach to address the first case where SAR backscatter is uncertain. In Fig. \ref{fig:dlr-fuzzy}a, the water/non-water separating threshold $\tau$ is defined as the upper fuzzy value $x_2$. This value represents the boundary between both classes, where the likelihood of a correct classification is the lowest. The mean backscatter value of the class water $\mu_{water}$ is associated with a minimum fuzzy value $x_1$. Pal and Rosenfeld (1988) \cite{Pal1988} describe the negative S-function that maps numeric to fuzzy values which is also depicted in Fig. \ref{fig:dlr-fuzzy}b.

As the majority of water pixels have backscatter values around the mean backscatter value, the uncertainty of a correct classification of these water pixels is low. The fuzzy logic approach maps high uncertainties to low degrees of membership to a particular class. For instance, high uncertainty of a correct pixel-wise classification to the water class corresponds to a low degree of membership to that class. The pixel is, therefore, assigned a lower fuzzy value. The converse is also true, where a low uncertainty corresponds to a high degree of membership of a given pixel to the water class; it is assigned a high fuzzy value. 

The three individual fuzzy membership functions are in the range [0, 1]. For easier interpretation and lower storage requirements, float values were rescaled to the range [0, 100]. The resulting fuzzy layer is computed as the mean of all three individual fuzzy layers. A defuzzification value of 60 is defined as the threshold to mark the distinction between water and non-water classes. 

Pixels classified as water with a fuzzy value of $\geq$ 60 are treated as water detections of high confidence.

\subsection{Computation of LIST probabilities}
\noindent The flood detection algorithm by LIST applies a change-detection approach \cite{Chini2017}, \ie \space the flood inundation is performed by detecting backscatter changes of two consecutive Sentinel-1 observations, the most recent SAR scene, $I_{t-1}^{S1}$ with the overlapping SAR scene acquired from the same orbit called reference SAR scene, $I_{t0}^{S1}$. As it is a change detection algorithm, it aims at detecting and mapping all decreases of backscattering values with respect to a reference one. A change detection approach is adopted because it allows to differentiate floodwater from permanent water bodies and, at the same time, filter out classes having water-like backscattering values such as shadows or smooth surfaces. The floodwater extent for the actual event is described with $I_{t0}^{S1}$ and the image difference to the pre-event situation is described with $I_D^{S1} = I_{t-1}^{S1} - I_{t0}^{S1}$. The likelihood of floodwater classification is characterized by flood probability. 

Both $I_{t0}^{S1}$ and $I_D^{S1}$ are used for likelihood estimation, the pixels that have high posterior probability of both water class and change class are likely to be real flooded pixels. The probability of being flooded for a given pixel (\ref{eq:list-1}) is defined as the minimum value between the conditional probability of the water class $p(W|\sigma^0)$ and the conditional probability of the changed class $p(C|\Delta\sigma^0)$ with regards to the Sentinel-1 backscatter $\sigma^0$:

\begin{equation}
\label{eq:list-1}
p(F|\sigma^0, \Delta\sigma^0) = min(p(W|\sigma^0), p(C|\Delta\sigma^0))
\end{equation}

where $p(\sigma^0)$ is the marginal distribution of backscatter values in $I_{t0}^{S1}$ and $p(\Delta\sigma^0)$ is the marginal distribution of backscatter difference values in $I_D^{S1}$ . In case a pixel is also flooded in the reference image, only $I_{t0}^{S1}$ is considered for likelihood estimation of flood classification, see (\ref{eq:list-2}).

\begin{equation}
\label{eq:list-2}
p(F|\sigma^0) = p(W|\sigma^0)
\end{equation}

As in this case, the likelihood is only calculated from the backscatter value in $I_{t0}^{S1}$, false high flood probability can be caused by permanent water and other water look-alike dark areas, these false alarms in binary map has been removed by comparing the resulting flood map with the previous flood map. To reduce these false high probabilities in current likelihood map, for non-flood pixels in the new flood map, their flood probability is the minimum value between $p(W|\sigma^0)$ and the value in the latest previous likelihood map.

\subsection{Computation of TUW uncertainties}
\noindent The flood algorithm by TUW is based on a data cube approach as introduced by \cite{BauerMarschallinger2022} and builds upon a-priori probability parameters for flood and non-flood conditions generated from Sentinel-1 time series. Incoming Sentinel-1 scenes that are subject to flood mapping are classified by means of Bayesian inference, which is not only computationally slim and NRT-suitable, but also intrinsically yields likelihood values in terms of posterior probabilities of the class allocation. For each pixel in a new Sentinel-1 backscatter measurement, the probability of belonging to either the flood or the non-flood class is inferred. 

Based on the Bayes decision rule, higher ("winning") posterior probabilities define then the class allocation. Additionally, the conditional error $p(error|\sigma^0)$ can be defined by the lower posterior probability, see (\ref{eq:tuw-1}), 

\begin{equation}
\label{eq:tuw-1}
p(error|\sigma^0) = min[p(F|sigma^0), p(NF|\sigma^0)]
\end{equation} 

where $p(F|sigma^0)$ describes the probability of the flood class and $p(NF|\sigma^0)$ the probability of the non-flood class with respect to the Sentinel-1 backscatter $\sigma^0$. 

The conditional error as direct measure for \textit{uncertainty} enables direct quantification of the lack of confidence with respect to a given decision. Since posterior probabilities sum up to 1, a higher posterior probability for one class results in a lower posterior probability for the other class in the binary classifications. Uncertainty is thus defined between 0.0 and 0.5. A value close to zero represents high confidence, since the probabilities for both classes (flood and non-flood) indicate a clear decision. High conditional errors (\ie \space close to 0.5) indicate uncertain decisions, as the new observation is falling into the overlap of the local flood/no-flood distributions and hence no class is much more probable than the other one. In such a situation, the Bayes decision is very uncertain and the classification is not meaningful. 

For all pixels of the incoming Sentinel-1 image, the conditional errors $p(error|\sigma^0)$ are forwarded to the ensemble algorithm, which represent the pixel-wise uncertainties associated with the flood map of TUW's algorithm. For easier interpretation and lower storage requirements, the uncertainties are scaled to values between 0 and 100. 

The TUW flood mapping algorithm features some internal masking of conditions not well represented by the a-priori probability parameters. This includes an internal uncertainty mask based on the statistical Sentinel-1 backscatter model of TUW is applied in this algorithm to exclude poorly-based decisions (\ie, with low reliability), defined by an upper limit of 0.2 for the conditional error, reflecting a 4:1 probability that
the assigned class is correct.

\subsection{Fusion of likelihoods}
\noindent In the context of this study, likelihood values of different origins are fused to a single quantity. Probability and fuzziness can be considered equal in terms of the numerical expression of the likelihood that is represented in the unit interval [0, 1]. However, they have to be differentiated in the manner in which the two measures handle the semantic classes water and non-water \cite{Kosko1990}. Given the same probability and fuzzy values of for example 0.8, the representation of likelihood is clarified as follows: 

A probability of 0.8 represents an 80 \% chance of pixel-wise water detection, where the value is determined based on pixel frequencies. The likelihood about the chance of a water or a non-water detection can be maximized as more observations become available and the pixel-wise water detection is built on a broader data base. 

A fuzzy value of 0.8 represents a pixel that is 80 \% water, describing the degree of membership belonging to that class, based on its properties. Such properties are defined through the uncertainty analysis, \eg \space the DLR algorithm attributes a pixel with Sentinel-1 backscatter, slope and size information that declare its membership to the class water. The fuzzy value expresses the degree to which it can be considered to be a (pure) water pixel. The uncertainty about class ambiguity persists even if more observations become available. Maximizing the likelihood is however possible by introducing additional auxiliary datasets that add to the pixel properties. 

This study acknowledges the mathematical and ontological complexities that characterize the formulation of the two aforementioned types of likelihood generation. However, in order to return actionable and interpretable information, the GFM likelihood product simplifies the fusion of likelihood values by computing the average value of the three algorithm likelihood outputs. While more advanced approaches have been proposed to bring the two measures of likelihoods together, \eg \space by \cite{He2015}, \cite{Kosko1990} and \cite{Hillenbrand2004}, this approach addresses the need for practicality in crisis information management. This objective is characterized by the need to make time-critical decisions with informative and also more easily interpretable products to support decision-making. Furthermore, the harmonized GFM likelihood product summarizes the likelihood inputs from the three water detection algorithms, thereby minimizing cognitive overload for end-users.

\section{Generation of ensemble likelihoods}
\noindent Combining the likelihood information generated by each flood algorithm requires value harmonization. TABLE \ref{tab:tab-1} outlines each of the three outputs with respective value ranges. The value ranges indicate the lowest and highest classification confidences as well as the threshold distinguishing flood from non-flood. Since the TUW algorithm outputs uncertainties, a pixel value of 100 represents a maximum uncertainty value that is comparable to a LIST probability or a DLR fuzzy value of 0. 

The DLR and LIST flood algorithms produce uncertainties that are numerically similar to likelihoods, with low values indicating low flood classification confidence and vice versa. The uncertainty analysis of TUW produces an inverse value range, where low values indicate high likelihood or high flood classification confidence and vice versa.

\begin{table}
	\begin{center}
		\caption{Likelihood quantifications of the three individual flood algorithms}
		\label{tab:tab-1}
		\begin{tabularx}{\columnwidth}{l l l}
			\hline
			Algorithm 	& Value ranges 				&
Type of 			\\
						& [low, threshold, high]	& likelihood		\\ \hline
			DLR			& [0, 60, 100]				& Fuzzy value		\\
			LIST		& [0, 50, 100]				& Probability		\\
			TUW			& [100, 50, 0]				& Uncertainty		\\ \hline
		\end{tabularx}
	\end{center}
\end{table}

By definition, a threshold value of 50 separates the flood and non-flood pixels in the ensemble likelihood layer. Fuzzy values F generated from the DLR algorithm are adapted to this scheme, based on (\ref{eq:dlr-1}). 

\begin{equation}
\label{eq:dlr-1}
likelihood_{DLR} = 
\begin{cases}
100 - 1.25 \cdot (100 - F), & F \geq 60 \\
\frac{F}{1.2}, & F < 60
\end{cases}
\end{equation}

The TUW uncertainties U are inverted, following (\ref{eq:tuw-2}).

\begin{equation}
\label{eq:tuw-2}
likelihood_{TUW} = 
\begin{cases}
100 - U, & flood_{TUW} = 1 \\
U, & flood_{TUW} = 0
\end{cases}
\end{equation}
		
Once likelihood values from all three algorithms are represented in the same range, the ensemble likelihood is computed as the mean of the individual likelihood layers, irrespective of nodata values. 

The ensemble algorithm computes the result on pixel level and requires two sets of three input layers each for flood and likelihood computations, respectively. If two flood algorithms fail to output data, the required flood and likelihood layers are generated automatically with the same geometry as the input Sentinel-1 scene and filled with zero values stating no flood for the entire scene, accompanied with zero values stating low likelihoods. 

A valid flood or a non-flood pixel is always connected to a valid likelihood pixel. If a flood pixel holds a nodata value, the corresponding likelihood pixel also stores a nodata value. 

This behavior has implications for the statistical robustness of the ensemble results. For instance, a flood pixel that is based on three valid individual classifications is considered to be statistically more robust compared to a flood pixel that is based on only two valid individual classifications (and one nodata classification). The latter is a so-called split situation that is resolved through a consensus approach, \ie, the ensemble algorithm marks that pixel as not flooded. 

\begin{figure}[!t]
	\centering
	\includegraphics{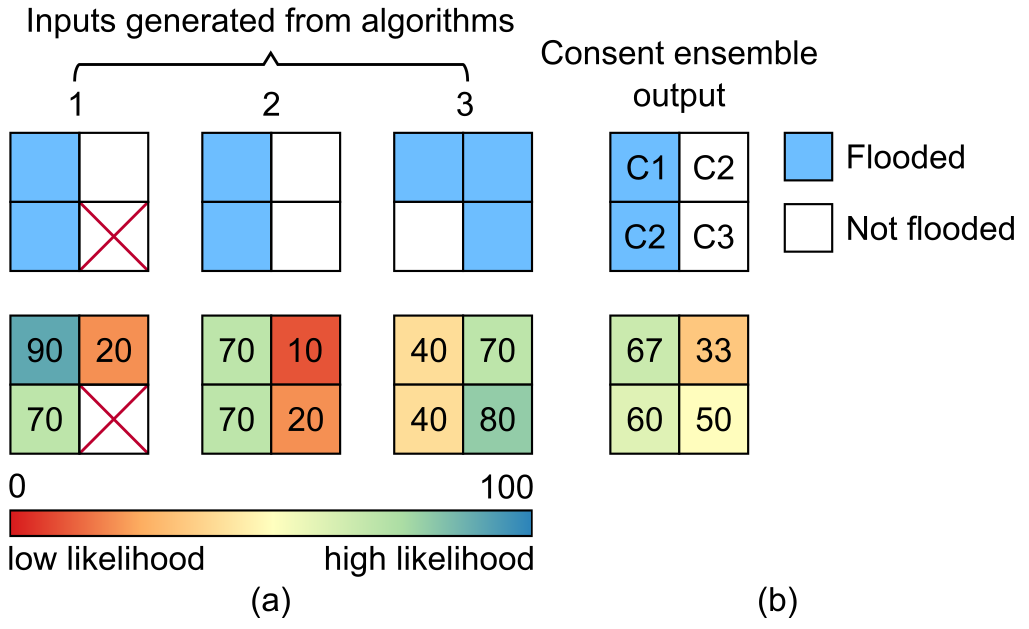}
	\caption{Sample inputs generated by each of the three individual flood algorithms with (a): available flood and likelihood data per algorithm on pixel level and (b): possible changes in the final classification. The cases C1 and C2 handle full and majority agreement respectively. The case C3 demonstrates a split situation that is resolved through a conservative agreement with a non-flood result.}
	\label{fig:likelihood-decisions}
\end{figure}

Fig. \ref{fig:likelihood-decisions} illustrates three different cases (C1, C2, C3) to demonstrate the ensemble classification scheme. In the first case C1, 3 out of 3 of the individual algorithms return a flood classification which is a full consent. In case C2, 2 out of 3 of the individual algorithms return a classification that disagrees with the third algorithm. This is a major consent and the pixel is classified as flood or non-flooded respectively, depending on the majority vote, \eg \space [flood, flood, non-flood] or [non-flood, non-flood, flood]. In case C3, one algorithm returns a nodata value and the remaining algorithms disagree on the classification, \eg \space [nodata, flood, non-flood]. The ensemble algorithm resolves the split situation through a conservative approach and marks the pixel as non-flooded. 

The steps described in this section define initial flood mapping likelihoods that are to be corrected with auxiliary data masking out error-prone regions and excluding areas of no interest, \eg \space reference water that is not flooded per definition. If these pixels are to be excluded and were classified as non-flooded, the initial likelihood value remains unchanged. If pixels to be excluded were classified as flooded, the respective likelihood value is changed to the value 49, \ie \space the most unconfident likelihood value for the non-flood class.

\section{Datasets}
\begin{figure}[!t]
	\centering
	\includegraphics{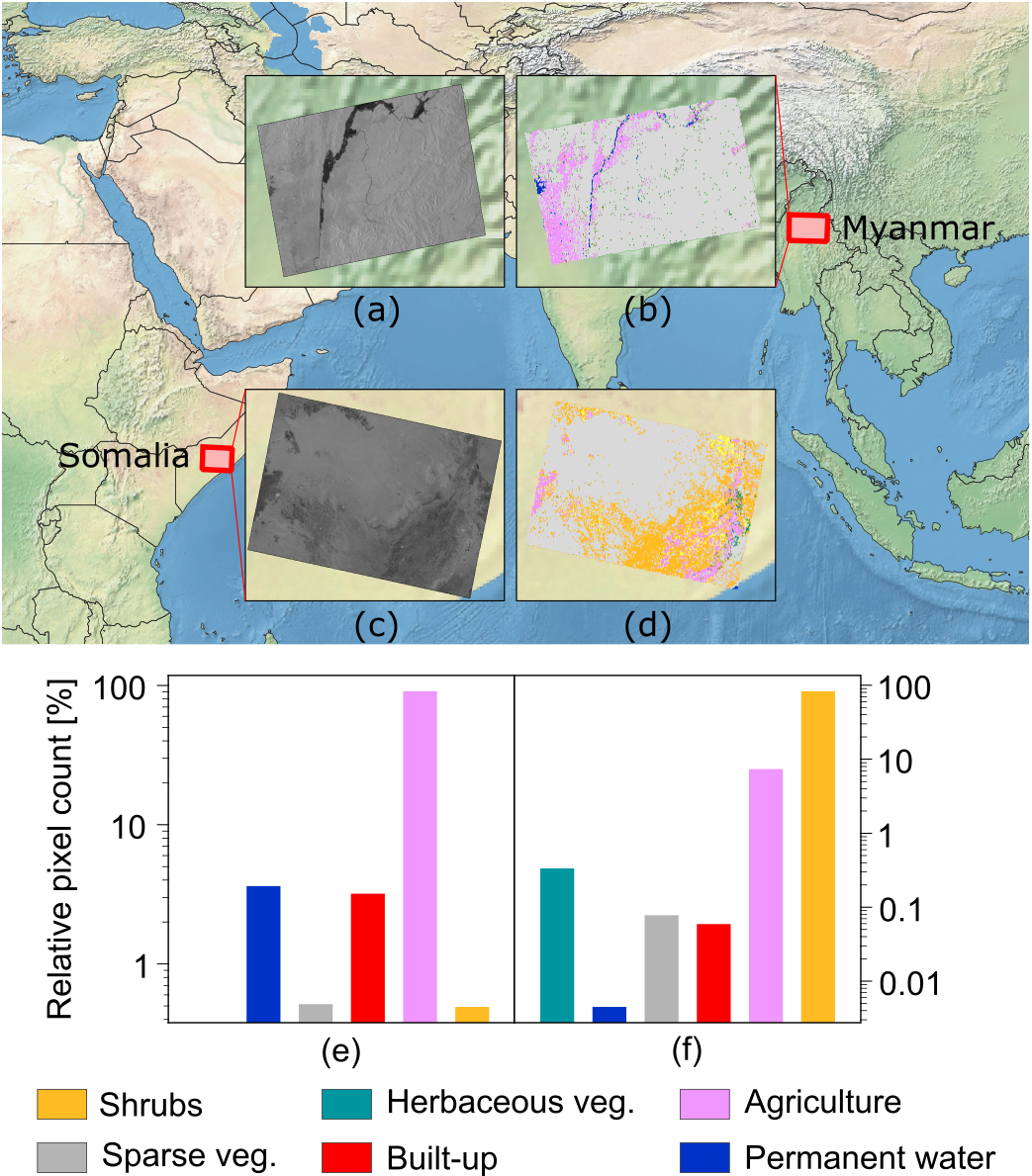}
	\caption{Location of the study areas in Myanmar and Somalia. The Sentinel-1 scene over Myanmar (a) was acquired on 201907-16 11:39:44. The Sentinel-1 scene over Somalia was acquired on 2019-03-16 02:46:06. Both study areas are accompanied with an overview of pre-dominant land cover maps (b) and (d). An exclusion mask is applied and shows areas where flood computation is performed. This information is in line with the pixel-wise distribution of land cover classes for Somalia (e) and Myanmar (f), valid for non-excluded regions.}
	\label{fig:datasets-lc}
\end{figure}

\noindent This section gives an overview on the used datasets and how they were processed within this study. Two use cases are presented that showcase a flood event in Myanmar and a non-flood situation in the semi-arid climate-zone of Somalia (see Fig. \ref{fig:datasets-lc}). The preprocessing of the Sentinel-1 IW GRDH datasets is described by the overview given in \cite{Wagner2021}.

The individual flood algorithms exploit Sentinel-1 GRD IW data which are shown in Fig. \ref{fig:datasets-lc} (a) and (c). Sentinel-1 data over Myanmar was acquired on 2019-07-16 11:39:44. Sentinel-1 data over Somalia was acquired on 2019-03-16 02:46:06. The fuzzy logic step of the DLR flood algorithm uses slope information derived from Copernicus DEM data \cite{CopDEM}. All input datasets were resampled to a common pixel spacing of 20 x 20 m in the Equi7Grid projection \cite{BauerMarschallinger2014}. 

Land cover information for this study is based on the global Copernicus Land Cover product from 2019 \cite{Buchhorn2020} with an original pixel spacing of 100 x 100 m that has been resampled to 20 x 20 m. As for this study, it was decided to focus on selected pre-dominant land cover types that are either of particular socio-economic interest or likely to be affected from flooding. Fig. \ref{fig:datasets-lc} (b) and (d) depict the spatial distribution of predominant land cover types for the study areas, followed by the class frequencies in Fig. \ref{fig:datasets-lc} (e) and (f). An exclusion mask is applied, leaving land cover types that are part of this analysis. 

For the Myanmar use case, the land cover type \textit{agriculture} dominates the study area with a pixel coverage of approx. 90 \% followed by the classes of \textit{built-up} and \textit{permanent water}, sharing less than 5 \% coverage each. Land cover types depicting forests and similar classes are not considered, although shown in green colors in the map, as they represent a state of dense vegetation that is mostly excluded from the flood computation.

For the Somalia use case, the land cover type \textit{shrubs} dominates the study area with a pixel coverage of approx. 90 \% followed by the class \textit{agriculture} with a pixel coverage of approx. 10 \%.

A crucial part of the presented data relies on a consistent and robust reference water dataset that was computed prior to the release of the GFM products. The GFM reference water dataset exploits a two-years' time series of Sentinel-1 median backscatter images that were aggregated for each month. Thus, the reference water reflects permanent water which is stable over the reference period of two years and seasonal water bodies that are periodically flooded over the duration of the reference period.

An exclusion layer defines pixels that are not included into the final ensemble output, as examined by \cite{Zhao2023submitted}. As already mentioned in section 2, the exclusion layer contains information about radar shadows, dense vegetation and permanent low backscatter, \ie, regions where the flood inundation is hampered, as well as topographic regions that are not prone to flooding.

In order to evaluate the ensemble likelihood values validation data covering the Myanmar study area is introduced. The data consists of a binary flood extent map derived from Sentinel-2 data, which was acquired on 2019-07-15 with a one-day delay to the acquisition of the Sentinel-1 data over Myanmar.

\section{Results}
\begin{figure}[!t]
	\centering
	\includegraphics{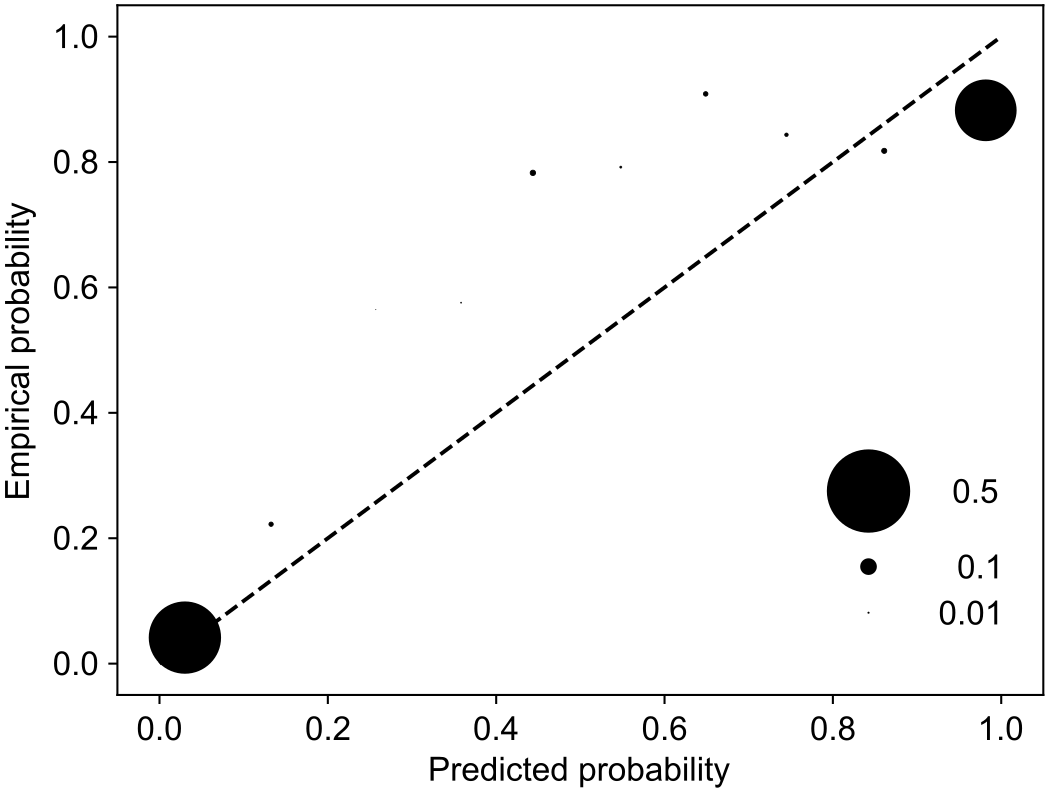}
	\caption{Quantile-quantile plot comparing predicted probabilities (ensemble likelihood values) with empirical probabilities from validation data. Each marker plots into the respective bin range, \eg \space [0.0, 0.1]. The marker size denotes the relative number of samples.}
	\label{fig:reliability-myanmar}
\end{figure}

\noindent In relation to the validation data, this study further examines a quantile-quantile plot supporting the evaluation of the ensemble likelihoods (see Fig. \ref{fig:reliability-myanmar}). The quantile-quantile plot shows the agreement of the computed ensemble likelihood values with the empirical probabilities. The majority ($>$ 80 \%) of the pixels fall into the first and last bins while the remaining pixels show higher empirical probabilities compared to the predicted samples.

To address the objectives defined in Section 1.2, scenarios are examined to identify specific likelihood regimes reflecting the number of algorithms that were used to compute the pixel-wise likelihood. 

\begin{figure*}[!t]
	\centering
	\includegraphics[width=\textwidth]{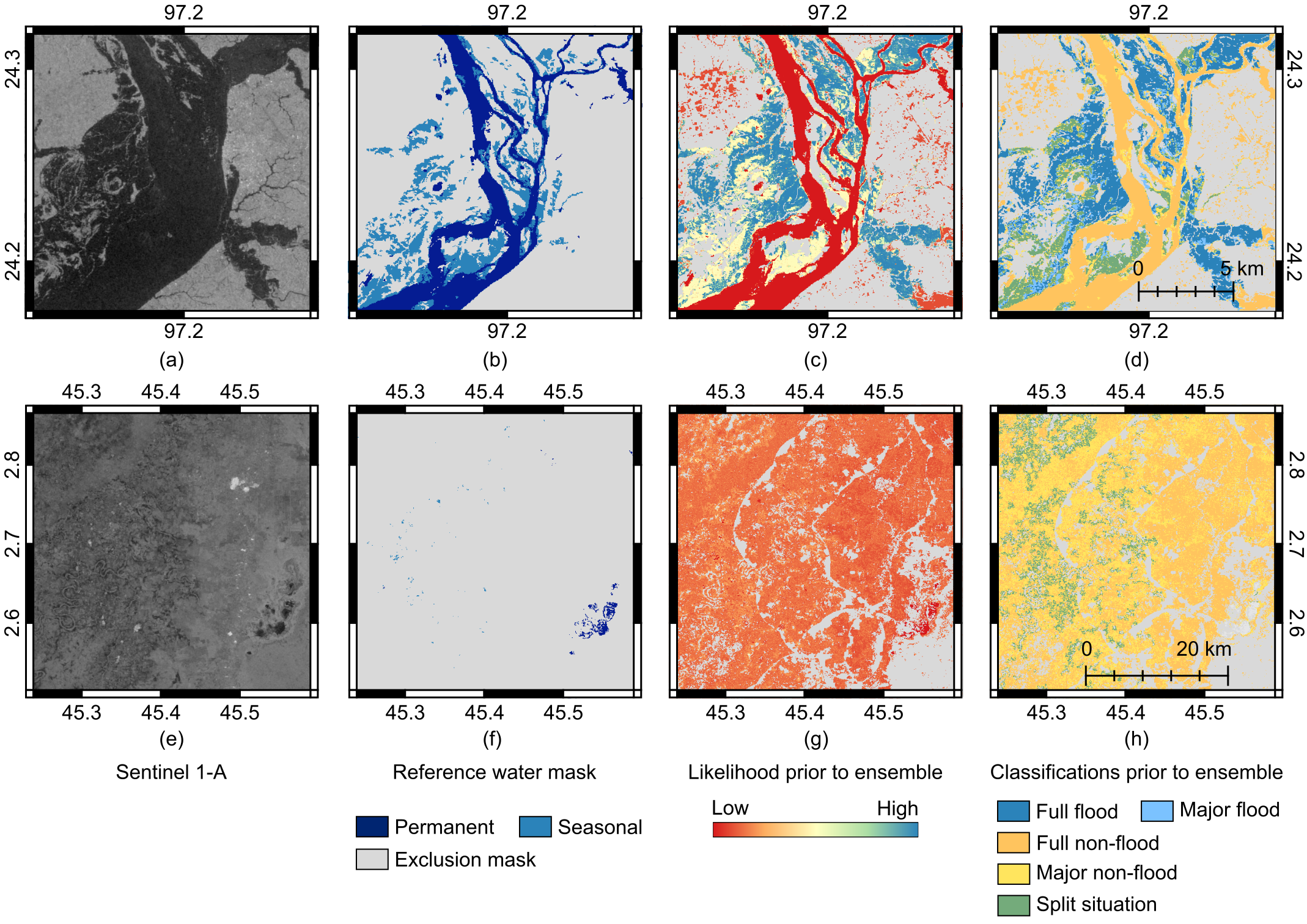}
	\caption{Series of image chips sampled from the Myanmar (a) (d) and Somalia (e) (h)use case: Sentinel-1A image with backscatter values (a) and (e), reference water and exclusion mask (b) and (f), likelihoods prior to ensemble (c) and (g) and spatial distribution of flood classifications prior to ensemble (d) and (h).}
	\label{fig:quadruplets}
\end{figure*}

The Myanmar use case represents a known flood event on July 16, 2019, see Fig. \ref{fig:quadruplets} (a) (d).

The likelihoods in subfigure (c) represent the initial mean likelihood values that were computed as the average of all available pixel-wise flood algorithm likelihood values, prior to the application of the ensemble algorithm. 

The classification subfigure (d) illustrates the initial flood classification prior to the application of the ensemble algorithm. Full consent marks pixels where 3 out of 3 algorithms agree on a flood and non-flood classification, respectively. Major flood and major non-flood indicate pixels where 2 out of 3 algorithms agree on the classification of flood or non-flood, respectively, \ie \space [flood, flood, non-flood], and vice versa with [non-flood, non-flood, flood]. It should be noted that [flood, flood, nodata] also results in a major flood decision; the same applies to major non-flood decisions with [non-flood, non-flood, nodata]. Split situations mark pixels where one algorithm cannot classify a pixel and thus outputs nodata, while the remaining algorithms disagree on the classification, \ie \space [flood, non-flood, nodata]. 

Very low initial mean likelihood values are observed over permanent water features (b). These values primarily correspond to areas that are excluded in the post-processing step and with initial full non-flood classifications (d). Much higher likelihood values are observed over image features that correspond to full or major flood classifications. Pixels with medium likelihood values around 50 correspond to split situations located along and within seasonal water bodies. The consensus approach resolves these split situations to non-flood decisions. 

\begin{figure}[!t]
	\centering
	\includegraphics{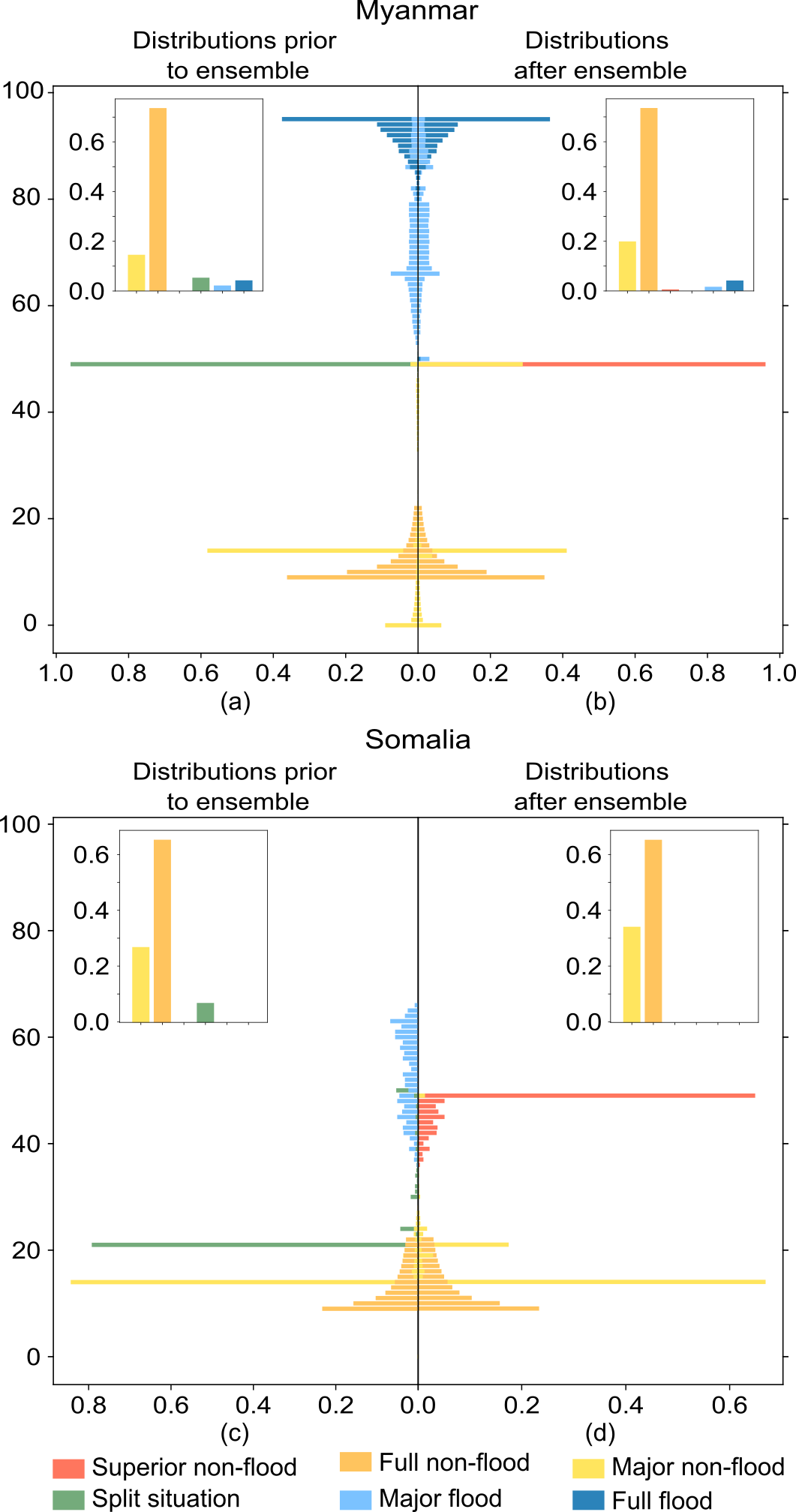}
	\caption{Histograms of likelihood distributions for the Myanmar (a) and (b) and Somalia (c) and (d) use case. Each box contains a histogram pair prior to the ensemble (a) and (c) and after the ensemble algorithm was applied (b) and (d). Pixels that were initially marked as flooded and have been overwritten by the exclusion layer to non-flooded are marked as superior non-flood. Depending on the distribution of the likelihood in the initial flood classification, ensemble flood likelihoods $<$ 50 can occur, \eg \space a classification with [flood, flood, non-flood] has the likelihoods [50, 50, 40] with a mean likelihood of 47. Each colored group sums up to 100 \%, \ie, the bar widths are not comparable but give an indication about the likelihood distribution within that group. The share of each group on the total pixel count is given with a bar plot for each of the histograms.}
	\label{fig:histograms}
\end{figure}

Fig. \ref{fig:histograms} illustrates split situations in green colors with likelihoods $<$ 50 that are remapped to non-flood decisions. Thus, 100 \% of the split pixels are re-classified to non-flood, thereby increasing the share of non-flood pixels for that particular likelihood value. Fig. \ref{fig:histograms}a also shows major flood pixels in the likelihood range [50, 80] and full flood pixels with likelihoods $>$ 80. A small number of major flood pixels with likelihoods $>$ 50 are excluded from the final ensemble results and therefore marked as superior non-flood with re-assigned likelihood values of 49. As can be seen in the top right bar plot of Fig. \ref{fig:histograms}b, the count of initial flood pixels that are remapped to non-flood pixels is very low, as indicated through the small size of the green bar. Also, the full non-flood class clearly dominates the classification types, followed by major non-flood decisions. In the Myanmar study area, about 10 \% of all pixels are classified as flooded. 

The Somalia use case represents a regular monitoring observation (\ie \space non-flood event) on March 16, 2019, where over-detections are likely to be observed as the environmental setting mostly covers dry soil, see Fig. \ref{fig:quadruplets} (e) (h). In effect, the use case contains a very limited number of water pixels in general, which is also reflected in the reference water mask (f). Very low initial mean likelihoods (g) highlight the fact that the majority of the pixels are initially classified as non-flooded (h). A substantial number of split situations are observed along meandering river channels. These split situations represent potential over-detections that appear as fragmented clusters and do not follow morphological shapes, \eg depression boundaries, and generally correspond to dark Sentinel-1 backscatter features, \ie \space potential water look-alikes. 

Fig. \ref{fig:histograms} (c) and (d) show medium to low likelihood values for split pixels that are remapped to major non-flood pixels and therefore increase the amount of major non-flood pixels, as depicted by the top right bar plot of (d). It is also clear that very few flood pixels are removed at the end of the ensemble algorithm with re-assigned likelihood values. Any initial flood pixel with a likelihood $>$ 50 is re-assigned with a likelihood value of 49 which is depicted in (d). However, the amount of these remapped pixels is still low and the majority of pixels belongs to non-flood classes. 

\begin{figure}[!t]
	\centering
	\includegraphics{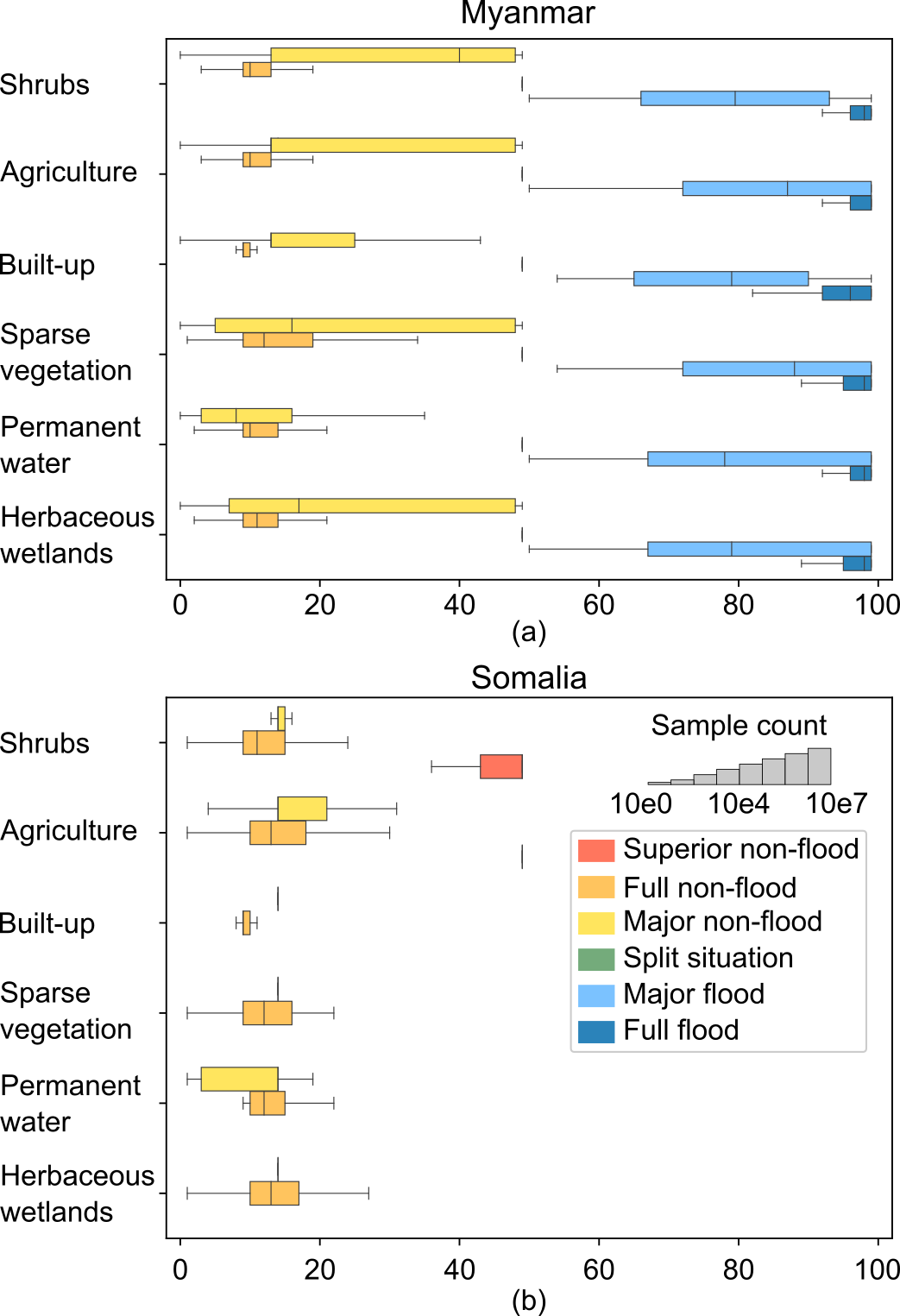}
	\caption{Boxplots of ensemble likelihood distributions with respect to land cover/ uses in the Myanmar (a) and Somalia (b) use case. The sample count for each box is indicated through the box height. Boxplots of extreme low variance are located along likelihoods of 50 and mark superior non-flood pixels. The legend in plot (b) applies for both plots.}
	\label{fig:lc-likelihoods}
\end{figure}

The next set of results supports the identification of land cover classes that are associated with different likelihood values (see Fig. \ref{fig:lc-likelihoods}). This analysis is performed with both use cases and aims to focus on land covers with relatively high economic and social impacts to end-users, \eg \space agriculture and built-up.

Fig. \ref{fig:histograms} already depicts low likelihood values for non-flood classifications for both use cases, which also dominate the amount of considered pixels for these classes. In congruence to that, Fig. \ref{fig:lc-likelihoods} shows low variance in likelihood values for full non-flood decisions across all land cover types and for both use cases. The situation differs for major non-flood decisions where the Myanmar use case shows greater likelihood variance across all land cover classes (Fig. \ref{fig:lc-likelihoods}a), in contrast to the Somalia use case (Fig. \ref{fig:lc-likelihoods}b). 

As stated in Section 4, the land cover type agriculture dominates the valid pixels in the Myanmar study area and attributes to the majority of full non-flood classifications that are depicted with likelihoods of 20 and less (Fig. \ref{fig:histograms}b). This is also shown in Fig. \ref{fig:lc-likelihoods}a where full non-flood classifications over the land cover type agriculture show very low variance. 

Major non-flood classifications show greater likelihood variance that originates from split situations, which are remapped to major non-flood in the ensemble algorithm. 

For the Myanmar use case, the less dominant flood pixels show low variance for full flood and greater variance for major flood decisions across all land cover types. Superior non-flood decisions are depicted with extremely low variance across all land cover types. 

As stated in Section 4, the land cover type shrubs dominates the Somalia study area and attributes to the majority of full non-flood classifications that are depicted with likelihoods of approx. 20, and less (Fig. \ref{fig:histograms}d). This is also shown in Fig. \ref{fig:lc-likelihoods}b, where full non-flood classifications over the land cover type shrubs show very low variance. 

Major non-flood classifications of greater variance are depicted for the land cover types agriculture, of which approx. 10 \% of pixels build the study area, and permanent water, that is almost not present in the region. However, both major non-flood clusters are rather small. 

As shown in Fig. \ref{fig:histograms}d and Fig. \ref{fig:lc-likelihoods}b, superior non-flood classifications occur and remap any potential flood classification prior to the application of the ensemble algorithm if they were masked out by the exclusion layer. Although only given with a very low number of pixels, their majority plots into the dominating land cover class shrubs and shows a likelihood cluster of low variances.

\section{Discussion}
\noindent The results provide a basis to obtain insights about the correlation of results prior to and after the application of the ensemble algorithm. In particular, the evaluations aim to link majority-based classifications and their statistical robustness to an explanatory variable, namely dominant land cover types.

\subsection{Quantile-quantile plot}
\noindent The quantile-quantile plot reflects the statistical reliability of the probabilistic prediction as the majority of predicted data is in close proximity to the one-to-one line. 

The part of the data showing high predicted probability marks a flood over-detection with reference to the validation data. At this point, it should be noted that the validation data is not real ground truth data but relies on the flood extent derived from Sentinel-2 data. Furthermore, the validation data shows a oneday delay to the acquisition of Sentinel-1 data, reflecting a probable change of the flood situation. Therefore, flood patches missing in the validation data can be considered as possible sources of flood over-detection in the prediction data. 

The parts of the data showing higher empirical probabilities with medium predicted probabilities mark flood underdetections with reference to the validation data. These regions mostly locate along the edges of detected flood patches and along river channels. It is likely that the exclusion layer does not cover these regions although they are likely to introduce misclassifications.

\subsection{Map subfigures}
\noindent Examining the series of subfigures for both use cases supports an initial assessment of the scenarios under which majority flood/non-flood classifications are identified, prior to the application of the ensemble algorithm. These results contain a significant number of split pixels where 1 out of 3 algorithms return a nodata value and 2 out of 3 disagree on the flood classification. This behavior does not indicate a failure of the system and is not to be mistaken as an inaccurate result. Nodata classifications occur if an algorithm is unable to classify a pixel with a robust likelihood, \ie \space if the result shows a significantly low classification confidence. Such an output is observed over challenging SAR image features that could not be excluded from the ensemble result. The ensemble algorithm translates these unconfident results to non-flood with a likelihood value of 49, \ie \space the most unconfident likelihood of the non-flood class. 

In the Myanmar flood event use case, the relatively high number of split situations coincide with the location of seasonal reference water bodies and with the presence of dense vegetation. Flood waters, including those classified based on majority decisions, also correspond with seasonal water bodies. Flood classifications over these areas represent a degree of disagreement among the contributing flood algorithms. The resulting mean likelihoods may be used to caution users to consider verifying these flood hotspots with additional data prior to making decisions \eg \space on resource allocation. Permanent water features, on the other hand, are classified as non-flooded despite 1 out of 3 algorithms classifying these features as flooded. It should be noted that the application of the reference water mask in the ensemble post-processing step reassigns the likelihood values of the permanent water features to 0 to indicate high confidence of being non-flooded. In these instances, the ensemble likelihood makes the non-flood classification explicit, regardless of the number of individually contributing flood algorithms over these pixels. 

In the Somalia use case, the relatively high number of split situations, in addition to major non-flood classifications, are located along former meandering river channels that seemed to have dried out and share the same SAR signal responses as bare soils. Herbaceous vegetation, shrubs and agriculture on dry soil are well-known challenges for flood detection based on Sentinel-1 backscatter, where the low radar backscatter tends to often result in over-detections. The consensus approach of the ensemble algorithm, although being a conservative measure, reduces these over-detections significantly and demonstrates the advantage of the ensemble algorithm over the application of a single measure alone.

\subsection{Histograms}
\noindent For the Myanmar use case, flood classifications are associated with ensemble likelihoods between 50 and 100; major floods detected with a certain degree of disagreement are associated with a wider range of lower ensemble likelihoods, while full floods are associated with a narrower range of higher confidence ensemble likelihoods between 80 and 100. The expression of agreement, as a form of confidence in flood detections, is useful information that can be consulted to support any decision making by end users during the onset of reported flood events. It should be noted that a major flood decision originates from 1 out of 3 algorithms classifying nodata or non-flood. Regardless of the scenario, such a likelihood value is rather low and clearly indicates lower confidence for the ensemble flood classification compared to a full flood agreement, which builds on a broader data basis. 

For the Somalia use case, no flood was classified, which is also a result of resolving split situations with the consensus approach. A different approach would have been to resolve split situations by favoring the flood or non-flood classification with the highest confidence in the respective class. Although less conservative, it would have been more likely to miss critical over-detections.

\subsection{Boxplots}
\noindent The pair of boxplots compare ensemble likelihood distributions with respect to land cover/uses in the Myanmar and Somalia use cases. 

In the Myanmar use case, ensemble likelihoods of flood detection increase from the lower-mid ranges corresponding to the initial mean likelihoods for all dominant land cover classes to notably higher ranges and demonstrate a medium spread of classification confidence. 

In the Somalia use case, no floods could be observed but in comparison to the Myanmar use case, the likelihood variances are rather low and demonstrate higher confidence of the nonflood classifications. 

At this stage, it should be noted that the exclusion layer also masks out regions that are likely to hamper the flood classification. It cannot be ruled out that for the Somalia use case, as a representative for challenging SAR-based flood detection conditions, a greater range of likelihood variance would be possible if another exclusion mask would have been applied to the data. However, this set demonstrates the advantage of including auxiliary data like an exclusion mask to focus on flood-prone regions. 

In comparison with the utilized land cover data, the permanent water class reveals regions of disagreement with the GFM flood product. Although the GFM flood and likelihood products are not relevant for permanent water features, an intersection of these datasets shows different results due to the significantly enhanced spatial resolution of 20 m in the GFM dataset compared to the spatial resolution of 100 m in the land cover dataset. 

Furthermore, the land cover class \textit{built-up} is to be excluded with the exclusion layer. However, the same diversity in spatial resolution applies to this case as well as the definition of built-up areas that are not meant to be included in the computation. Apart from a small number of towns, both study areas also contain light settlements that are not covered by the exclusion mask. Having the likelihood values for these built-ups is useful information for the end-users as this land cover type is of high socio-economic interest.

\section{Conclusion}
\noindent Within this paper, we describe a methodology to combine flood ensemble likelihoods of the Sentinel-1 GFM product. We further highlight the importance of interpretable and robust likelihood values to guide end-users and decision-makers in their processes. 

The computation of likelihoods informs on the robustness of flood classifications. While various methods have been proposed in the literature to combine different types of uncertainty information origins, \eg \space probabilities and fuzzy values, their computation and fusion is rather complex and arguably hampers their interpretability and a straight-forward crisis response. In contrast, the method presented here is easy to interpret and its application is straight forward, as it is solely based on the computation of the arithmetic mean of the individual flood algorithm likelihoods. Considering the value range, the classification of a flood pixel with an associated likelihood close to 100 is considered to be more confident than a flood pixel with a likelihood of close to 50, and is based on a wider range of input data. Furthermore, flood classifications with low likelihood values, \ie \space values in the range [50, 60], originate from an ensemble configuration with one algorithm classifying a pixel as non-flooded or nodata. Consequently, the ensemble likelihood product alerts end-users to the presence of non-consent flood classifications, which should be treated with care in decision making. 

The first results show how ensemble likelihoods function as a heuristic to identify and provide a first indication of the performance, as well as the agreement among the three algorithms contributing to the final flood classification. These aggregated likelihood values capture cumulative uncertainties from data, model architecture, algorithmic level and interpretation. Further reduction of uncertainties requires more dedicated investigative methods. Once generated, this kind of uncertainty information can be used by both of the aforementioned communities. In particular, researchers or algorithm developers are offered guidance to investigate how to minimize uncertainties with respect to certain explanatory variables, \eg \space land cover types. End users may consult likelihood information that supports cautioning against the direct use of flood classification product in areas with low likelihoods of flood classification, or where classifications are based on majority, rather than full detections. For resource allocation, it may be sufficient to identify areas of certain extents as potential hotspots, even if likelihoods associated with individual pixels in the vicinity of the areas of interest are lower. 

Based on the preliminary results, the benefits and limitations of the ensemble likelihood approach are highlighted and provide a starting point for further developments and applications in the two research and end-user communities. Further assessments may be conducted to include additional variables, in addition to extending the number and variety of use cases.

\printbibliography

\begin{IEEEbiography}[{\includegraphics
		[width=1in,height=1.25in,clip,
		keepaspectratio]{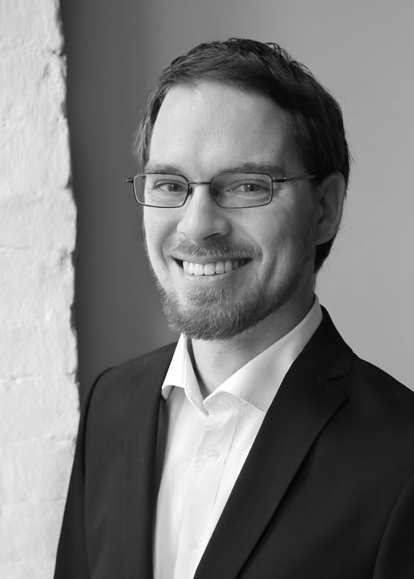}}]
	{Christian Krullikowski}
	 received the BSc degree in Geological Sciences at the Freie Universität Berlin, Germany, in 2013 and his MSc degree in Geological Sciences with a specialization in Hydrogeology at the Freie Universität Berlin, Germany, in 2016. In 2017, he joined the German Remote Sensing Data Center (DFD) of the German Aerospace Center (DLR) in the department Geo-Risks and Civil Security. His research activities focus on thematic processing with rule-based and machine learning methods for efficient disaster response.
\end{IEEEbiography}

\begin{IEEEbiography}[{\includegraphics
		[width=1in,height=1.25in,clip,
		keepaspectratio]{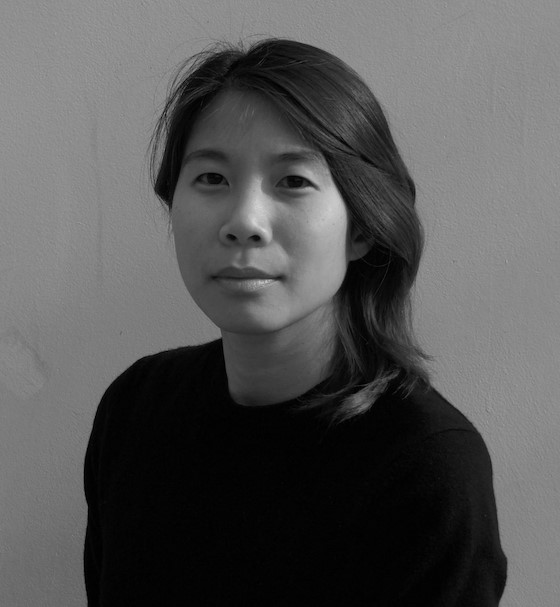}}]
	{Candace Chow}
	completed studies at the Universität Bern and participated in an Erasmus Mundus program specializing in flood risk management. In 2020, she joined the German Remote Sensing Data Center (DFD) of the German Aerospace Center (DLR) in the Geo-Risks and Civil Security department. She worked at the interface of research and operations, focusing on uncertainty quantification and communicating time-sensitive information to end users, particularly within the context of the International Charter "Space and Major Disasters".
\end{IEEEbiography}

\begin{IEEEbiography}[{\includegraphics
		[width=1in,height=1.25in,clip,
		keepaspectratio]{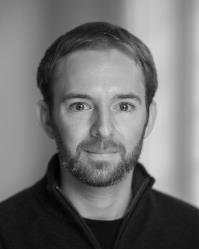}}]
	{Marc Wieland}
	received the Diploma degree in Geography from the RuprechtKarls Universität Heidelberg, Germany, in 2009 and the PhD degree from the Technical University of Berlin in 2013. In 2010 he joined the German Research Centre for Geoscience in Potsdam. In 2015 he moved to Chiba University, Japan to work on statistical pattern recognition in SAR timeseries. In 2016 he joined the University of Oxford as postdoctoral researcher. He is currently based at the German Remote Sensing Data Center (DFD) of the German Aerospace Center (DLR) where his research activities focus on machine learning techniques for emergency response.
\end{IEEEbiography}

\begin{IEEEbiography}[{\includegraphics
		[width=1in,height=1.25in,clip,
		keepaspectratio]{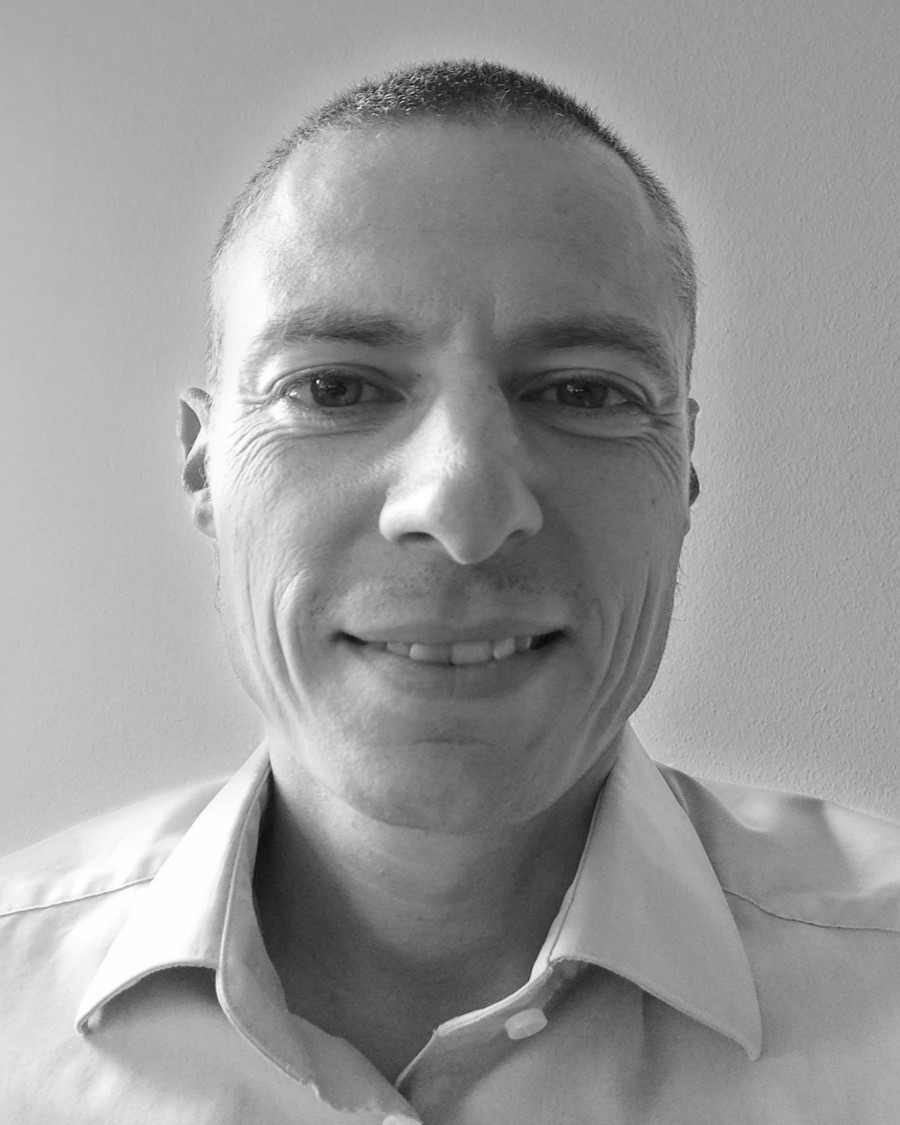}}]
	{Sandro Martinis}
	received the Diploma degree in Geography, Physics, and Remote Sensing from the University of Munich, Germany, in 2006. He received the PhD degree from the University of Munich in 2010 working on automatic flood detection using high resolution X-band SAR satellite data at DLR. From 2006-2007, he was a research associate with the University of Munich working on the development of remote sensing-based methods for the monitoring of glacier motions and subglacial volcanic eruptions. Since 2013, he is head of the research group "Natural Hazards" within the department Georisks and Civil Security of DLR. Since 2016 he is leading the operational activities of Germany's contribution to the International Charter "Space and Major Disasters".
\end{IEEEbiography}

\begin{IEEEbiography}[{\includegraphics
		[width=1in,height=1.25in,clip,
		keepaspectratio]{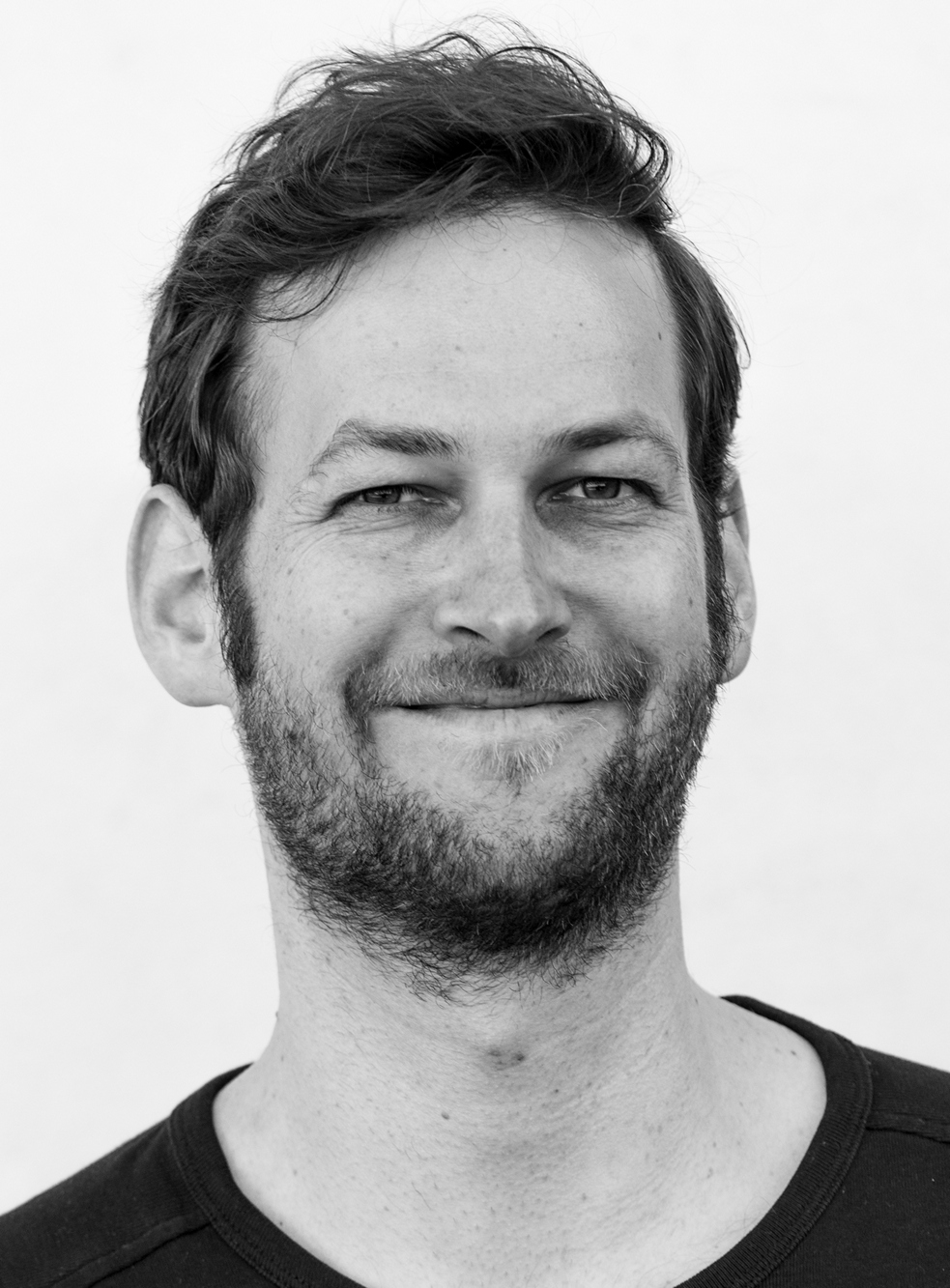}}]
	{Bernhard Bauer-Marschallinger}
	received his doctoral degree in Geodesy and Geoinformation from the TU Wien in 2018. As part of the TU Wien Remote Sensing group of the Department of Geodesy and Geoinformation, he has been involved i.a. in projects from ESA, EU, Copernicus Global Land Services (CGLS), Copernicus Emergency Management Services (CEMS), and Austrian Research Promotion Agency (FFG). His focus is on the exploitation of satellite data and the development of operational processing software for Synthetic Aperture Radar (SAR) observations, aiming for algorithms that operate globally and in near-real-time. With his team, he pursues the enhancement of the Sentinel-1 SAR processing and of scalable and efficient spatial reference systems. His scientific interest is in the retrieval of geophysical variables like soil moisture, water bodies and flood dynamics, employing signal and image analysis methods driven by mathematics and physics.
\end{IEEEbiography}

\begin{IEEEbiography}[{\includegraphics
		[width=1in,height=1.25in,clip,
		keepaspectratio]{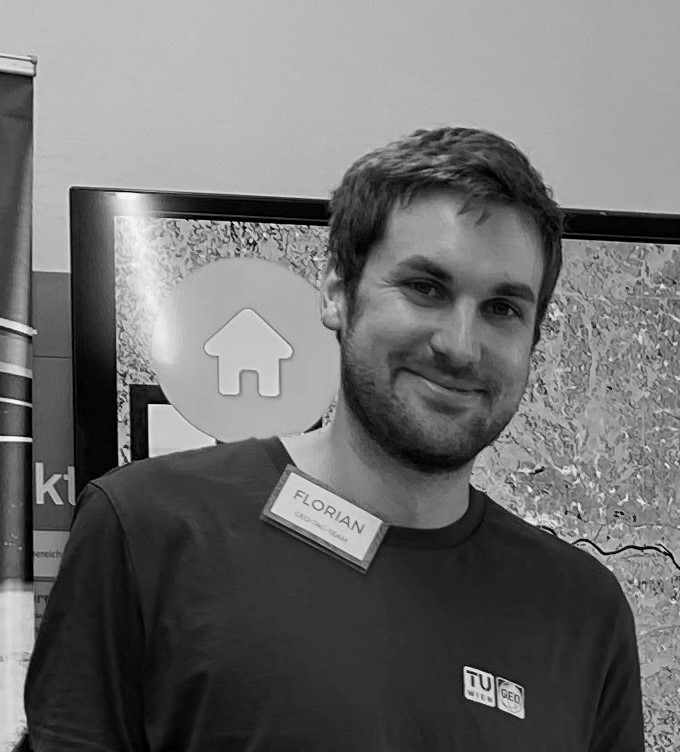}}]
	{Florian Roth}
	completed his Bachelor and Master studies in Geodesy and Geoinformation at the Technische Universität Wien (TU Wien) while putting the focus on remote sensing and geoinformation. He joined the Remote Sensing group of TU Wien's Department of Geodesy and Geoinformation in 2020. His research interest lies in retrieving and evaluating flood information from SAR-based satellite data.
\end{IEEEbiography}

\begin{IEEEbiography}[{\includegraphics
		[width=1in,height=1.25in,clip,
		keepaspectratio]{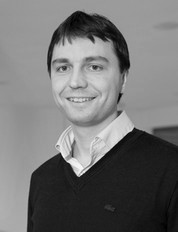}}]
	{Patrick Matgen}
	earned his M.Sc. in Environmental Engineering at the Ecole Polytechnique Fédérale de Lausanne in 2002. In 2011 he finalized his PhD dissertation at the Technical University in Delft. Since January 2020 he is leading the group "remote sensing and natural resources modelling" at the the Luxembourg Institute of Science and Technology. The group's aim is to develop the synergistic use, processing and interpretation of data from multiple complementary active and passive sensors installed on both space and airborne platforms and to provide deeper insights into the relations between spectral features and properties of Earth's natural resources. His personal research interests include the assimilation of remote sensing-derived observations into hydrodynamic models and the assessment of flood hazard and risk at large scale.
\end{IEEEbiography}

\begin{IEEEbiography}[{\includegraphics
		[width=1in,height=1.25in,clip,
		keepaspectratio]{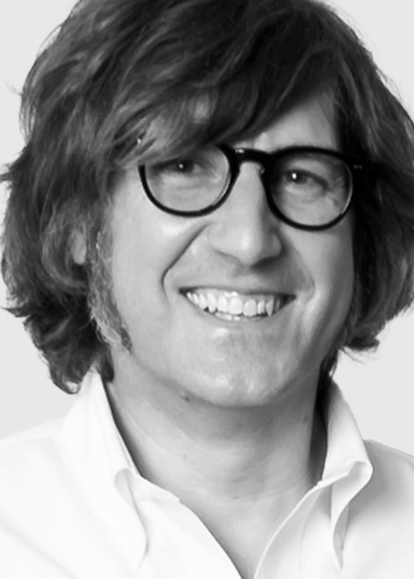}}]
	{Marco Chini (M’09–SM’11)}
	earned his M.Sc. in Environmental Engineering at the Ecole Polytechnique Fédérale de Lausanne in 2002. In 2011 he finalized his PhD dissertation at the Technical University in Delft. Since January 2020 he is leading the group "remote sensing and natural resources modelling" at the the Luxembourg Institute of Science and Technology. The group's aim is to develop the synergistic use, processing and interpretation of data from multiple complementary active and passive sensors installed on both space and airborne platforms and to provide deeper insights into the relations between spectral features and properties of Earth's natural resources. His personal research interests include the assimilation of remote sensing-derived observations into hydrodynamic models and the assessment of flood hazard and risk at large scale.
\end{IEEEbiography}

\begin{IEEEbiography}[{\includegraphics
		[width=1in,height=1.25in,clip,
		keepaspectratio]{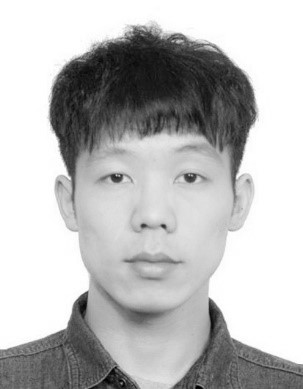}}]
	{Yu Li}
	received the BSc degree in geoinformation science and technology and the MSc degree in mathematical geology from China University of Geosciences (Wuhan), in 2012 and 2015, respectively, and the PhD degree in physical geography from LMU Munich, Germany, in 2020. From 2016 to 2019, he worked as a PhD student at the German Aerospace Center (DLR), Oberpfaffenhofen, Germany. He is currently a Junior Research and Technology Associate at the "Remote Sensing and Natural Resources Modeling" group, Luxembourg Institute of Science and Technology (LIST), his research interests include machine/deep learning, multi-temporal image analysis, and natural disaster monitoring with SAR data.
\end{IEEEbiography}

\begin{IEEEbiography}[{\includegraphics
		[width=1in,height=1.25in,clip,
		keepaspectratio]{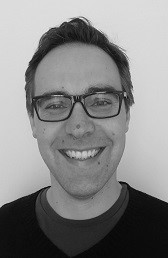}}]
	{Peter Salamon}
	is a scientific project manager at the Joint Research Center of the European Commission where he is coordinating the Copernicus Emergency Management Service and its components the European and Global Flood Awareness Systems. He holds an M.Sc. in Applied Environmental Geoscience from the Eberhard-Karls University in Tübingen (Germany) and a PhD in Hydraulic and Environmental Engineering from the Polytechnic University of Valencia (Spain). During his research activities and professional career, he has gained experience in policy support in the area of disaster risk management at European and global level. His technical background is in hydrology, numerical modeling, hazard and risk mapping, climate change impacts, uncertainty assessment and operational flood forecasting. He is author and co-author of more than 50 peer-reviewed articles.
\end{IEEEbiography}

\end{document}